\documentclass[12pt,preprint]{emulateapj}

\usepackage{graphicx,amsmath,natbib}

\slugcomment{To be submitted to PASP}

\shorttitle{A New High Contrast Imaging Program at Palomar Observatory}
\shortauthors{Hinkley et al.}

\begin{document}

\title{A New High Contrast Imaging Program at Palomar Observatory}

\author{Sasha Hinkley            \altaffilmark{1,10}}
\author{Ben R. Oppenheimer\altaffilmark{2}} 
\author{Neil Zimmerman        \altaffilmark{3,2}}           
\author{Douglas Brenner       \altaffilmark{2}}
\author{Ian R. Parry \altaffilmark{4}}                               
\author{Justin R. Crepp  \altaffilmark{1}}                       
\author{Gautam Vasisht            \altaffilmark{7}} 
\author{Edgar Ligon                  \altaffilmark{7}}
\author{David King                  \altaffilmark{4}}              
\author{R\'emi Soummer        \altaffilmark{5}}               
\author{Anand Sivaramakrishnan\altaffilmark{5,2,6}}   
\author{Charles Beichman       \altaffilmark{9}}
\author{Michael Shao              \altaffilmark{7}}            
\author{Lewis C. Roberts, Jr.  \altaffilmark{7}}            
\author{Antonin Bouchez         \altaffilmark{8}}
\author{Richard Dekany           \altaffilmark{8}}
\author{Laurent Pueyo              \altaffilmark{7}}
\author{Jennifer E. Roberts     \altaffilmark{7}}
\author{Thomas Lockhart        \altaffilmark{7}}              
\author{Chengxing Zhai           \altaffilmark{7}}    
\author{Chris Shelton             \altaffilmark{7}}          
\author{Rick Burruss                 \altaffilmark{7}}

\altaffiltext{1}{Department of Astronomy, California Institute of Technology, 1200 E. California Blvd, MC 249-17, Pasadena, CA 91125}
\altaffiltext{2}{Astrophysics Department, American Museum of Natural History, Central Park West at 79th Street, New York, NY 10024}
\altaffiltext{3}{Department of Astronomy, Columbia University, 550 West 120th Street, New York, NY  10027}
\altaffiltext{4}{Institute of Astronomy, Cambridge CB3 0HA, United Kingdom}
\altaffiltext{5}{Space Telescope Science Institute, 3700 San Martin Drive, Baltimore, MD 21218}
\altaffiltext{6}{Stony Brook University}
\altaffiltext{7}{Jet Propulsion Laboratory, California Institute of Technology, 4800 Oak Grove Dr., Pasadena CA 91109}
\altaffiltext{8}{Caltech Optical Observatories, California Institute of Technology, Pasadena, CA 91125}
\altaffiltext{9}{NASA Exoplanet Science Institute, California Institute of Technology, Pasadena, CA 91125}
\altaffiltext{10}{Sagan Fellow}

\begin{abstract}
We describe a new instrument that forms the core of a long-term high contrast imaging program at the 200-in Hale Telescope at Palomar Observatory.  The primary scientific thrust is to obtain images and low-resolution spectroscopy of brown dwarfs and young Jovian mass exoplanets in the vicinity of stars within 50 pc of the Sun. The instrument is a microlens-based integral field spectrograph integrated with a diffraction limited, apodized-pupil Lyot coronagraph. The entire combination is mounted behind the Palomar adaptive optics system. The spectrograph obtains imaging in 23 channels across the $J$ and $H$ bands (1.06 - 1.78 $\mu$m).  The image plane of our spectrograph is subdivided by a $200\times200$ element microlens array with a plate scale of 19.2 miliarcsec per microlens, critically sampling the diffraction-limited point spread function at 1.06 $\mu$m. In addition to obtaining spectra, this wavelength resolution allows suppression of the chromatically dependent speckle noise, which we describe.  In addition, we have recently installed a novel internal wave front calibration system that will provide continuous  updates to the AO system every 0.5 - 1.0 minutes by sensing the wave front within the coronagraph.  The Palomar AO system is undergoing an upgrade to a much higher-order AO system (``PALM-3000"): a 3388-actuator tweeter deformable mirror working together with the existing 241-actuator mirror.  This system, the highest resolution AO corrector of its kind, will allow correction with subapertures as small as 8.1cm at the telescope pupil using natural guide stars.  The coronagraph alone has achieved an initial dynamic range in the $H$-band of $2\times10^{-4}$ at $1^{\prime\prime}$, without speckle noise suppression.  We demonstrate that spectral speckle suppression is providing a factor of 10-20 improvement over this bringing our current contrast at $1^{\prime\prime}$ to $\sim$$2\times10^{-5}$.  This system is the first of a new generation of apodized pupil coronagraphs combined with high-order adaptive optics and integral field spectrographs (e.g. GPI, SPHERE, HiCIAO), and we anticipate this instrument will make a lasting contribution to high contrast imaging in the Northern Hemisphere for years.  
\end{abstract}


\section{Introduction}
In recent years, astronomers have identified more than 400 planets outside our solar system, launching the new and thriving field of exoplanetary science \citep{mul08,mbf05}.  
The vast majority of these objects have been discovered indirectly by observing the variations induced in their host star's light.  Radial velocity surveys can  provide orbital eccentricity, semi-major axes, and lower limits on the masses of companion planets while observations of transiting planets \citep{dsr05, sba08, cbi09} can provide fundamental data on planet radii and limited spectroscopy of the planets themselves.  
However, studying those objects out of reach to the radial velocity and doppler methods, will more fully probe the global parameter space occupied by exoplanets \citep{oh09}. The technique of direct imaging is a promising method for detecting and, more importantly, studying in detail the population of wide separation exoplanets \citep[see e.g.,][]{bkt10}.   Moreover, recent results \citep{mmb08,kgc08,lbc10} have demonstrated that direct imaging of planetary mass companions and disks \citep{obh08,hos09,mss09,bab09} is a technique that is mature and may become routine using ground-based observatories. 

Direct imaging surveys \citep[e.g.][]{mmh05,bcm07,ldm07,ncb08,chv09,clb10,lsh10,lwb10} hold the promise of quickly settling some of the most basic puzzles about exoplanetary systems, such as the true fraction of stars hosting planets, the dependence of this fraction on stellar environment, and a more robust measure of multiplicity in exoplanetary systems.  Future surveys can also quickly assess the distribution of planets beyond 5-10 AU.  The orbital placement of these companions will enlighten ongoing work into planetary migration, and help to constrain models of planet formation, evolution, and dynamical histories.  Once the instrumentation and techniques are fully in place, direct detection will not only be an extremely efficient method of discovery \citep[see e.g.][]{tcj09,zoh10}, but also a powerful tool to probe the nature of exoplanets. 

The ultimate goal of direct imaging surveys is not just to examine planet orbital parameters and the related implications for formation scenarios.  Direct imaging enables spectroscopy \citep[see e.g.][]{jbg10,bld10}, the key to unlocking the detailed properties of the objects themselves.  Spectroscopy provides clues to the atmospheric chemistry, internal physics, geology, and perhaps even sheds light on astrobiological activity associated with these objects \citep{kc03,ksf10}.  More robust classification schemes for planets in general will arise from observing as many planets as possible at different ages, in different environments, and with a broad range of parent stars.

\section{Challenges to High Contrast Imaging}
The major obstacle to the direct detection of planetary companions of nearby stars is the overwhelming brightness of the host star.  If our solar system were viewed from 20 pc, Jupiter would appear $10^8-10^{10}$  times fainter than our Sun in the near-IR \citep{bcb03,b05} at a separation of 0.25$^{\prime\prime}$, completely lost in its glare. The key requirement is the suppression of the star's overwhelming brightness through precise starlight control \citep{oh09,am10}. 
 
\begin{deluxetable}{ll}
\tabletypesize{\scriptsize}
\tablecaption{Overview of Project 1640 characteristics}
\tablewidth{0pt}
\tablehead{ \colhead{}  & \colhead{}}
\startdata                                     
Telescope:                                                                                                                       &                                            \\
                            \hspace{8pt}Aperture (m)                                                                 & 5.1                                      \\
                            \hspace{8pt}$\lambda/D$ at 1.06 $\mu$m (mas)                      & 43                              \\
                            \hspace{8pt}Telescope output f/ratio                                             & 15.4                                        \\ 
                            & \\
AO System:										&\\
			\hspace{8pt} Deformable Mirrors ($N_{\rm act}$) & 241 and 3388 actuators  \\
			\hspace{8pt} Subaperture size (cm)			& 63.6, 32.4, 16.2 and 8.1  \\
                            \hspace{8pt} Max. AO Control radius (mas)\footnotemark[1]  & 1010 ($\lambda$=1.06$\mu$m)-- \\                            										                          &1818 ($\lambda$=1.78 $\mu$m) \\
										                                                &                                              \\
Coronagraph:                                                                                                                    &                                                \\
                           \hspace{8pt}Apodizer diameter (mm)                                    &  3.80 \\
                           										& 2\% undersized from pupil                                      \\
                           \hspace{8pt}Astrometric Grid                                     &  $\Delta {\rm m} = 7.4$ \\
                           							& 4 spots at 22$\lambda/D$                                     \\
             
                            \hspace{8pt}Focal Plane Mask Size                                             &  370 mas  \\
                                                                                                                                        & =5.37 $\lambda/D$ at $\lambda=$1.65$\mu$m \\
                                                                                                                                         &=1322 $\mu$m for f/149.1 beam \\
		        \hspace{8pt}Undersized Lyot Stop Factor                                     &  2\% from Apodizer                                       \\
                                                                                                                                             & 4\% from primary pupil                  \\
                           \hspace{8pt} Final f/ratio	& f/164.6 at IFS microlenses\\
                           & \\

IFS:                                                                                                                                      &                                                \\
                            \hspace{8pt}Wavelength Coverage ($\mu$m)                           &   1.06 - 1.78                \\ 
                            \hspace{8pt}IFS Field of View (mas)                                             &    3840                           \\ 
                            \hspace{8pt}IFS pixel scale (mas/microlens)                               &   19.2                   \\ 
                           \hspace{8pt}Microlens Pitch ($\mu$m)                                          &   75                            \\ 
                            \hspace{8pt}Number of Spectra                                                               &   $200\times200 = 4\times 10^4$ \\ 
                            \hspace{8pt}Spectral Resolution ($\lambda/\Delta\lambda$)  &   33 to 58                                             \\ 

\enddata
\footnotetext[1]{AO control radius is defined as $N_{\rm act}\lambda/2D$ and discussed in e.g. \citet{osm03}}
\label{overviewtable}
\end{deluxetable}

A promising method for direct imaging of stellar companions involves two techniques working in conjunction.  The first, high-order Adaptive Optics (hereafter ``AO''), provides control and manipulation of the image by correcting the aberrations in the incoming stellar wave front caused by the Earth's atmosphere.  Second, a Lyot coronagraph \citep{l39,skm01} suppresses this corrected light.  Together, these two techniques can obtain contrast levels of $10^4$-$10^5$ at 1$^{\prime\prime}$.  Improvements in coronagraphy, specifically the apodization of the telescope pupil \citep{asf02,saf03,s05} can significantly improve the achieved contrast, especially at high Strehl ratios.

\begin{figure}[ht]
\center
\resizebox{1.1\hsize}{!}{\includegraphics[angle=-90.]{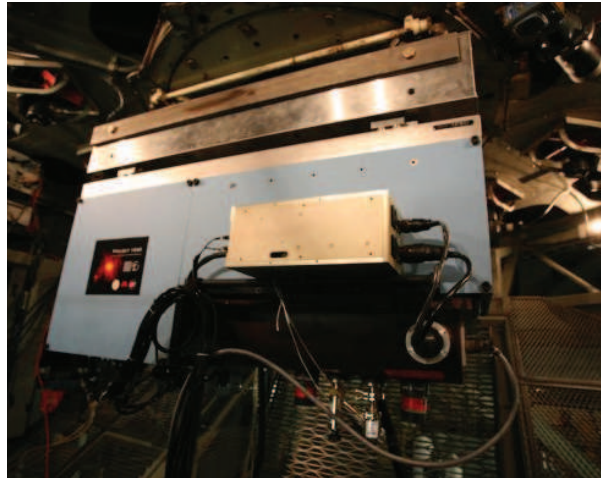}}
  \caption{The Project 1640 IFS, coronagraph and precision wave front calibration system mounted on the Palomar Adaptive Optics system.  The whole assembly is mounted at the Cassegrain focus of the 200-in Hale Telescope.}
  \label{p1640photo} 
\end{figure}

The combination of techniques described above, coronagraphy and high-order adaptive optics shows promise for direct imaging.  But this combination of techniques, like any form of high contrast imaging, still suffers from a significant source of residual noise, limiting the detection sensitivity.  Small phase aberrations in the incoming stellar wave front, arising from imperfections in the AO optics or the coronagraphic optics, can lead to a pattern of speckles that litter the image in the focal plane \citep[e.g.][]{l95,bdt01,brr02,psm03,as04} .  These speckles have lifetimes of hundreds of seconds or longer \citep[e.g.][]{hos07}, and hence are the single largest hindrance to the detection of faint companions around nearby stars. Without a coronagraph, \citet{rwn99} has demonstrated that speckle noise can dominate over photon shot noise by a factor of $\sim$$10^4$.  Such speckle noise is largely due to non-common path errors (those not measured by the AO wave front sensor), e.g., small aberrations in the coronagraphic optics, as small as 1 nm occurring ``downstream'' of the wave front sensor. These optical aberrations translate directly to errors in the incoming stellar wave front. Speckle noise is also highly correlated and thus not surmountable with simple techniques such as long exposures, or using larger and larger telescopes \citep{slh02,hos07,sfa07}.

\begin{figure}[ht]
\center
\resizebox{1.0\hsize}{!}{\includegraphics[angle=0.]{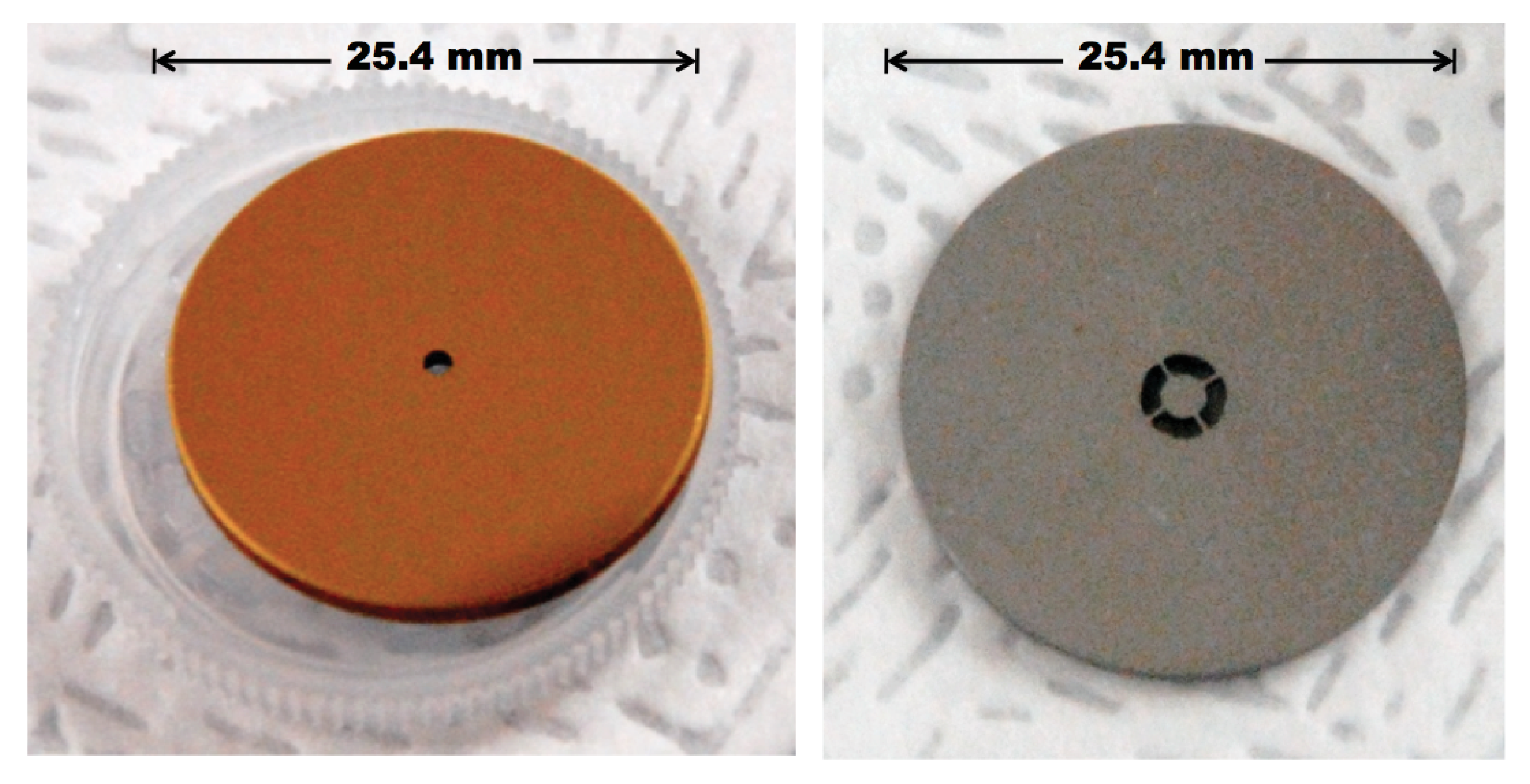}}
  \caption{{\it Left:} an image of the 5.37$\lambda/D$ occulting mask. Rather than an opaque mask, the Project 1640 coronagraph uses a reflective surface with a hole serving as the primary occulting disk.  The light captured through the hole is used as a reference arm for the precision wave front calibration system (see Section~\ref{calsystem}.) The light not entering the hole is reflected on to OAP3.  {\it Right:} an image of the Lyot mask, of which the transmissive portion has been undersized by 4\% from the telescope primary mirror. 20\% of the light passing through the Lyot mask is used as the ``science arm'' for the wave front calibration system, while the remaining 80\% is passed onto the IFS.  }
  \label{masks} 
\end{figure}

The suppression of speckle noise is paramount in the task of direct detection of exoplanets. Several past studies have explored two such methods of speckle suppression.  One method involves simply subtracting speckles that are highly stable in time through image post-processing \citep[e.g.][]{mld06, bcl04}.  This method can easily improve the  detection sensitivity by at least one to two magnitudes \citep{lsh10}, a factor of a few to ten.  Another method uses dual-imaging polarimetry \citep{kpp01,obh08, hos09}.  This technique is extremely powerful, essentially eliminating the unpolarized speckle pattern and greatly increasing sensitivity to polarized objects, such as circumstellar disks \citep{gkm07,obh08,pvg09}, achieving a contrast of nearly $10^{-4}$ (10 mag) at 0.4$^{\prime\prime}$ separations \citep[see e.g.][]{hos09}.  

Another promising speckle suppression technique can eliminate speckle noise but without the limitation that the target of observation be polarized.  The speckle noise pattern is an optical phenomenon, and its morphology is wavelength dependent. Indeed, the position of each speckle will move radially outward from the image center with increasing wavelength.  This wavelength dependence allows differentiation between the speckle noise and true astrophysical objects. This method, sometimes referred to as ``spectral deconvolution'' has been described before, e.g. \citet{sf02, lmd07}.  An integral field spectrograph is well suited to take advantage of the wavelength diversity shown by the speckle noise pattern.  Integral field spectroscopy has been used in the past for high contrast imaging science \citep{mml07,tat07,jbh08}, and we have built a customized instrument specifically dedicated to this task.

\begin{figure}[ht]
\center
\resizebox{1.0\hsize}{!}{\includegraphics[angle=0.]{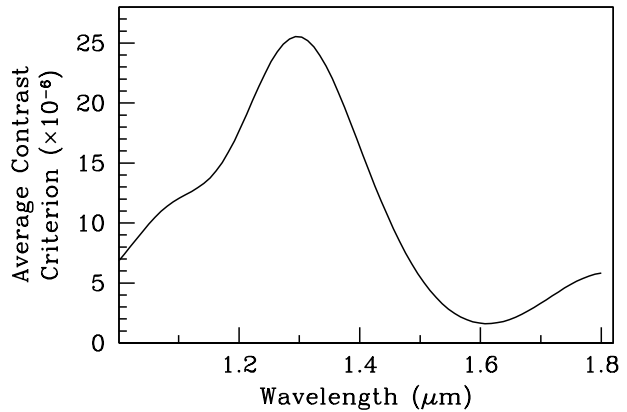}}
  \caption{Coronagraphic optimization of the average contrast outside the inner working angle and within the control region of the AO system, as a function of wavelength.  The optimization of the coronagraph includes the apodization function, the focal plane mask diameter and the Lyot stop geometry. This metric is dominated by the contrast right outside the inner working angle. The actual contrast improves markedly with increasing separation.}
  \label{optimization} 
\end{figure}

Our new project, dubbed ``Project 1640'' \citep{hob08} incorporates all of these advances with an integral field spectrograph being coupled to an Apodized-pupil Lyot Coronagraph, a precision wave front calibration system, and integrated to a high-order AO system which we describe below. The instrument mounted on the Palomar AO system is shown in Figure~\ref{p1640photo}.  Although we have chosen not to incorporate any polarimetric capabilities (See Table~\ref{overviewtable}), the overall design parameters for this project are otherwise similar to future high contrast surveys like The Gemini Planet Imager \citep{mgp08}, SPHERE \citep{bfd08} or HiCIAO \citep{mks08,mg09}.

\section{Coronagraph}\label{coronagraphsec}
To suppress the starlight of our target stars, we have built an apodized-pupil Lyot coronagraph (APLC) \citep[][Soummer et al. 2010]{sl05,s05}, an improvement of the classical Lyot coronagraph \citep{l39,skm01}.  We achieve the majority of our suppression with a reflective focal plane mask with a 1322$\mu$m diameter hole shown in Figure~\ref{masks}. We use the hole as an opaque mask and let the unocculted portion of the image around the hole be reflected on to the rest of the optical train.

\begin{figure}[ht]
\center
\resizebox{1.1\hsize}{!}{\includegraphics[angle=0.]{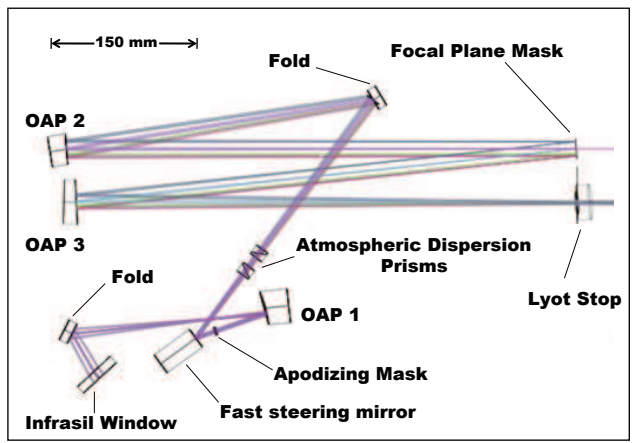}}
  \caption{The layout of our Apodized pupil Lyot Coronagraph (see text) taken from a Zemax design.  The coronagraph design is based on \citet{s05,sl05} and Soummer et al. 2010 (submitted).  
 The system also has two sets of rotating prisms to correct for differential atmospheric refraction. Details of the various masks are given in the text and Table~\ref{overviewtable}. All optics shown in this figure lie in the same plane of the coronagraph optical bench. A final spherical mirror (not shown) sits out of the plane of the optics in this figure and delivers the beam reflected from the Lyot stop to the Integral Field Spectrograph.}
  \label{coronagraph} 
\end{figure}

The IFS operates from 1.06 to 1.78$\mu$m, and a single coronagraph design is used for this entire range of wavelengths. This  very wide bandpass (covering both $J$ and $H$ bands) makes the coronagraph optimization challenging. The apodization function, focal plane mask size and Lyot stop have all been optimized for maximal starlight suppression over the $J$ and $H$ bands simultaneously, using the approach described in Soummer et al. 2010 (Submitted). The APLC coronagraph set (apodization function, focal plane mask diameter, and Lyot stop geometry) was optimized using  the average contrast in the AO controlled region of the focal plane, and outside of the inner working angle (IWA) as the optimization criterion. The coronagraphic suppression is unconstrained in the gap between the $J$ and $H$ bands. A plot of this optimization criterion is shown in Figure~\ref{optimization}.  The best theoretical contrast for the Project 1640 APLC design is achieved in the $H$-band, while allowing for good contrast in the $J$-band. This choice was in part motivated by the lower performance of the AO system at shorter wavelengths ($J$ band) therefore not requiring as much coronagraph starlight suppression.

\subsection{Coronagraphic Layout}
\begin{figure}[ht]
\center
\resizebox{1.0\hsize}{!}{\includegraphics[angle=0.]{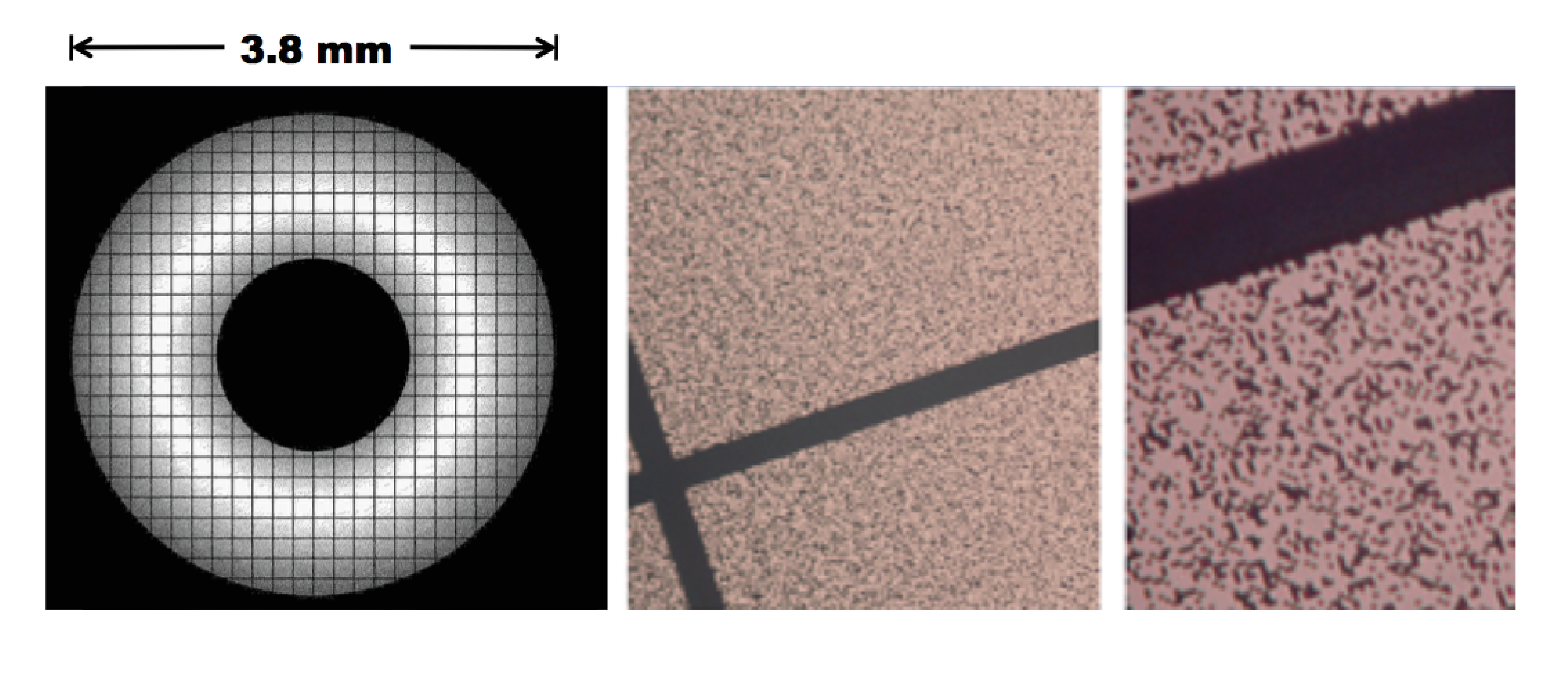}}
  \caption{{\it Left:} The transmission profile of the apodizing mask placed in the first pupil plane in the coronagraph.  The mask is 3.8 mm in size and is undersized from the telescope pupil by 2\%.  The superimposed grid used to create fiducial reference spots for astrometry can be seen over the transmission profile.  {\it Middle and right:} magnifications of the mask showing the results of the ion-beam etching process to generate the 10 $\mu$m-sized microdots which form the transmission profile.}
  \label{apodizer} 
\end{figure}

The layout of the APLC is shown in Figure~\ref{coronagraph} and is similar to a classical Lyot coronagraph, using an additional pupil apodization. The f/15.4 beam from the Palomar AO system enters our coronagraph via an infrasil window which counteracts dispersion caused by the PALAO dichroic. The beam comes to a focus, then comes out of focus, strikes an Off-Axis Parabola (``OAP 1'' in Figure~\ref{coronagraph}),  which forms a collimated beam. Next in the optical train is the transmissive apodizing mask built by Jenoptik and shown in Figure \ref{apodizer}, which is imprinted on infrasil glass using ion-beam etched microdots each 10 $\mu$m in size \citep{ssc09,sso09,sso10}.  The density of dots varies to produce the apodization function.  The apodizer is also imprinted with a grid that produces fiducial reference images of the star 20 $\lambda/D$ away from the central star.  These images are 7.4 magnitudes fainter than the primary star and are used for astrometric measurements (better than 2 mas rms error) while the star is occulted by the focal plane mask \citep{so06,mlm06}.

The beam then strikes the fast-steering mirror and continues onto a pair of atmospheric dispersion prisms (more detail on these is given below). The beam is brought back into a focus by a second powered optic, ``OAP 2,'' in order to apply the primary coronagraphic suppression at the focal plane mask. The 1322 $\mu$m diameter of the hole in the focal plane mask which serves as the occulter, corresponds to 5.37$\lambda/D$ at $1.65 \mu$m.
The light that has passed through the hole, then passes through the wave front calibration system (Section~\ref{calsystem}), and is reimaged onto an infrared Hamamatsu InGaAs quad-cell sensor, sensitive from 1.0 to 1.7 $\mu$m. This quad-cell sensor serves to keep the star centered on the occulting spot, and adjustments to the quad-cell position in turn will adjust the position of the star relative to the occulting spot.  The center of the stellar image is maintained on the sensors using a centroiding algorithm in conjunction with a PID control loop which drives the fast-steering mirror (FSM, a Physik Instrumente S-330.30 piezo-electric tip/tilt platform). A similar setup using optical APDs is described in \citet{dho06}. 
Performing the tip/tilt control in the same wavelength range as our science wavelength ensures that the coronagraph is optimally performing at the science wavelength.  Our fast steering mirror is updated at a ~1 kHz frequency to maintain the position of the star on the center of the spot, and we achieve a residual image motion of less than 5 mas per minute. We are able to operate this tip/tilt system on stars as faint as magnitude 6, and magnitude 7 under very good conditions.

The unocculted portion of the image is reflected off the focal plane mask, and travels to OAP 3 which re-collimates the beam.  Immediately prior to a pupil image, the beam is split, with 20\% of the light passing into the wave front calibration system (see section~\ref{calsystem} below).  The other 80\% of the light is reflected and forms a pupil image at the Lyot stop. The Lyot stop is a wire-EDM cut steel disk 0.25 mm thick, coated with Epner LaserBlack shown in Figure~\ref{masks}.  The Lyot stop suppresses the bright outer regions of the pupil image, the telescope secondary mirror and spider structures, and passes the rest of the image on to the rest of the system. The Lyot stop outer diameter is undersized by 2\% with respect to the apodizer diameter based on mechanical alignment tolerances. The Lyot stop inner diameter is increased by 25\% compared to the apodizer central obstruction according to the result of the optimization for broadband chromatic starlight suppression. After the Lyot stop, the beam travels to a final 600 mm spherical mirror which forms an image on the lenslet array of the spectrograph (not shown in Figure~\ref{coronagraph}). The entrance beam into the spectrograph has an f-ratio of 164.6 including the aperture downsizing from the pupil plane stops.

Due to differential atmospheric refraction \citep[e.g.][]{r02}, at an angle of 50$^\circ$ off of zenith, the angular position of a star at 1.06 $\mu$m and that at 1.78 $\mu$m will differ by $\sim$175 mas, essentially the radius of our focal plane mask. At such an angle, this means that observations at 1.06 $\mu$m will show the star to be completely occulted, while at 1.78 $\mu$m the star may be on the edge of the coronagraphic mask, largely unocculted. In order to perform effective coronagraphy across such a broad spread in wavelength, at zenith angles greater than $\sim$30--50$^\circ$, compensation must be implemented to account for this dispersion.  The Atmospheric Dispersion Correcting prisms are comprised of two sets of of Risley prisms, each one formed by two wedges of BaF$_2$ and CaF$_2$, the tips of which have been cemented together \citep[e.g.][]{w96,w97}. A cylinder has been cored out of each cemented wedge pair and mounted into a motorized, rotating mount.  Each cemented set can rotate relative to each pair or in tandem together to correct for the differential atmospheric refraction caused by the Earth's atmosphere. 

\section{Integral Field Spectrograph Design}
We have built a microlens-based integral field spectrograph (``IFS'') covering 1.06---1.78 $\mu$m with a 4$^{\prime\prime}$ field of view. This will allow us to obtain low-resolution spectra ($R\sim$ 33--58) at all $4\times10^4$ image samplings.

\begin{figure}[ht]
\center
\resizebox{1.05\hsize}{!}{\includegraphics[angle=0.]{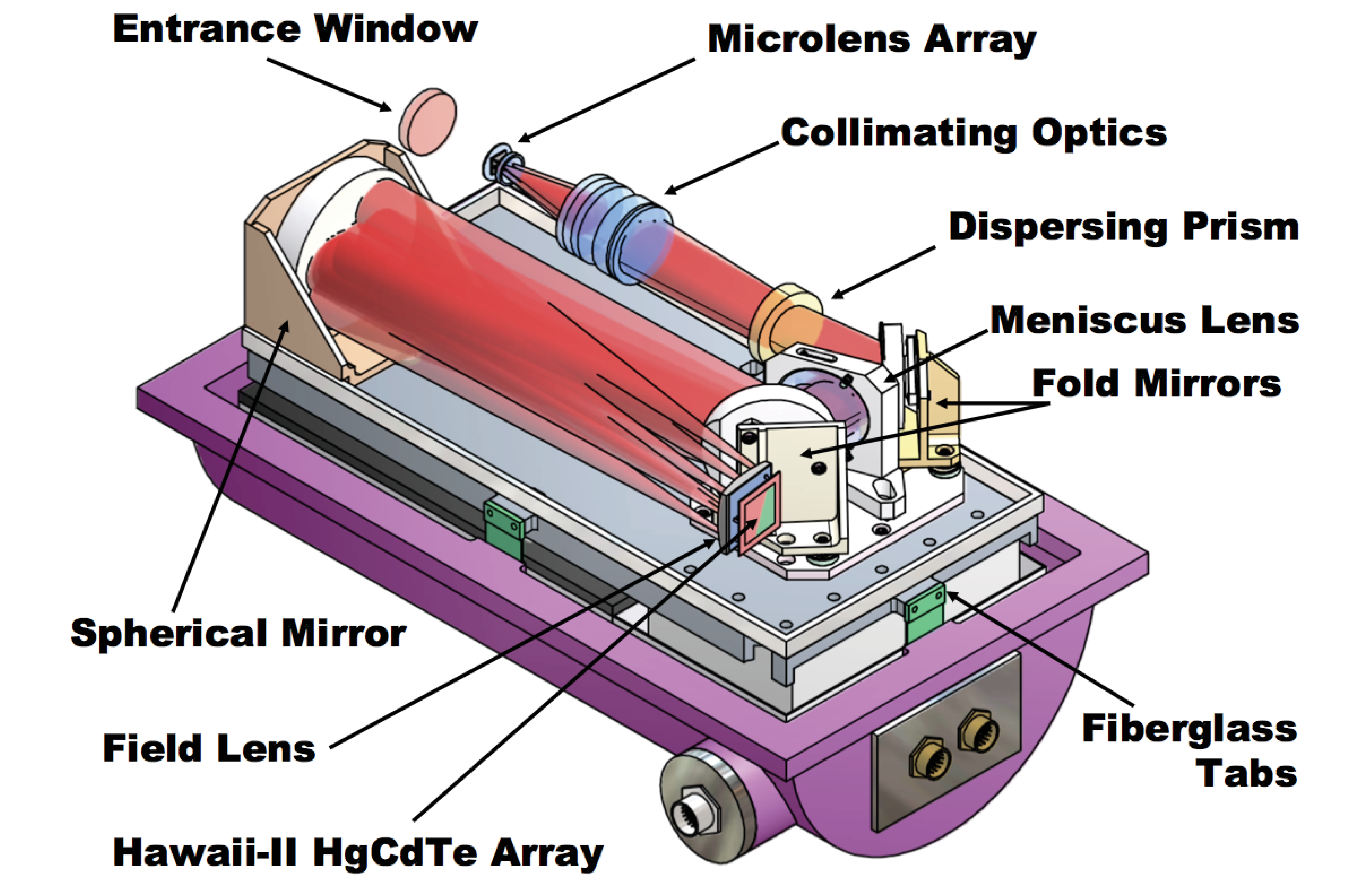}}
  \caption{Details of the internals of the Project 1640 Integral Field Spectrograph.  In this CAD drawing, the dewar lid and aluminum heat shield have been removed to reveal the internal optics of the IFS, and the optical layout. 
  }
  \label{dewarinternals} 
\end{figure}

\subsection{Optical Design}
The optical design for our integral field spectrograph is shown in Figure~\ref{dewarinternals} below, and is similar to the TIGER-type microlens-based IFS \citep{bab95,mpv08}.  The overall design can be categorized into four components: 1) a microlens array; 2) a dioptric collimator with a 160 mm focal length consisting of five lenses, which forms a pupil image on the input face of the disperser; 3) a prism/disperser; and 4) a 400 mm focal length catadioptric camera, which produces the spectral images on the detector.  Our transmissive optics were manufactured by Janos Technology, except for the microlens array, which was made by MEMS Optical.  The reflective optics were manufactured by Axsys technologies. We discuss each of these components in more detail below. 

\begin{figure}[ht]
\center
\resizebox{.97\hsize}{!}{\includegraphics[angle=0.]{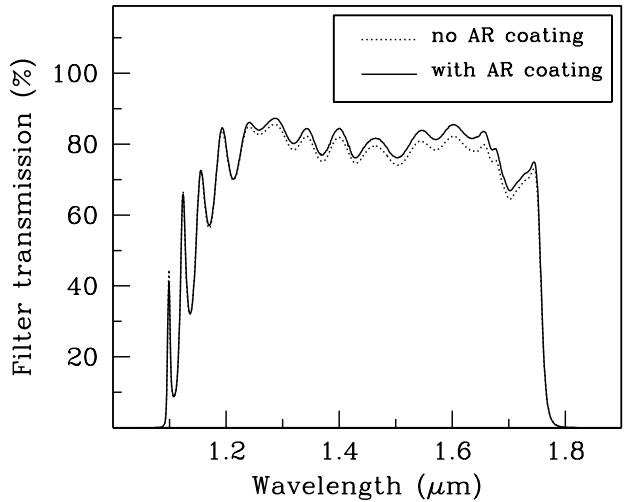}}
  \caption{The transmission profile for the blocking filter located in the IFS, directly in front of the Rockwell HAWAII-2 HgCdTe infrared detector.  Two curves are shown: one measuring the transmission prior to receiving the anti-reflective coating, and the transmission after.}
  \label{filter} 
\end{figure}

The optical design has been fully modeled using the Zemax design software, including the effects of thermal contraction as the system is cooled.  The system does not perform at non-cryogenic temperatures. All optics with the exception of the microlens array are oriented square with the optics base plate.  To prevent our spectra from overlapping on the detector, the orientation of the microlens array is rotated by $18.43^{\circ}$ from the normal vector to the base plate. The prism is oriented to disperse the light parallel to the base plate.  This places the detector square with the base plate as well, requiring only a rotation on the lenslet array. Wavelength filtering is achieved with $J$ and $H$-band blocking filter (1.06 - 1.78 $\mu$m), with OD3-OD4 blocking outside this range, placed directly in front of the detector. The transmission profile for the blocking filter prior to, and after receiving an anti-reflective coating is shown in Figure~\ref{filter}.

The square microlens array consists of two powered faces etched into a 1 mm thick wafer of fused silica.  The microlenses on the incident face have a radius of curvature of 950 $\mu$m and is primarily used to prevent cross-talk between microlenses, so that the higher powered exit surface retains as much of the light as possible (with minimal loss due to roll-over between the microlenses).  The rear face microlenses have a 159 $\mu$m radius of curvature to create the pupil images 280 $\mu$m behind the microlens array substrate.  A similar design is being incorporated into the SPHERE and GPI project IFSs \citep[e.g.][]{mgp08, bfd08, adg09}.  Each microlens has a pitch of 75 $\mu$m.  The effective f-number of each microlens is f/4, measured using the diagonal of each square microlens (106.1 $\mu$m).  We have $270\times270$ microlenses on our array, but only $200\times200$ of them are used. The array is mounted 4 mm directly in front of the first lens of the collimator. 

The assembly containing the collimator consists of five lenses mounted in a single housing. The lens materials are BaF$_2$, SF2, and SK8. The collimator forms 40 mm pupil images at the incident surface of the prism.  The prism is a single piece of BK7 glass, with a wedge angle of $4^{\circ}$ on each face, 60 mm in diameter. This prism is optimized for the wavelength range (1.06 --- 1.78 $\mu$m) with a dispersion direction parallel with the plane of the base plate.

The camera optics within the IFS refocus the image onto the detector.  These consist of a meniscus corrector lens, spherical mirror, and a field-flattening lens (field lens) in front of the detector as shown in Figure~\ref{dewarinternals}.  Both lenses are fused silica. The meniscus corrector lens has two surfaces with slightly different radii of curvature (216.82 and 227.90 mm), and was cored out of a larger (240 mm diameter) parent lens, 80 mm off-axis. The final field lens creates a flat focal plane for the final detector.  This lens is incorporated into the mount holding the detector and was custom designed to have adjustment capabilities in three-dimensions.  This lens serves the double purpose of providing additional protection for the detector.  
We also utilize two fold mirrors to accommodate packaging. All mirrors are made of diamond turned aluminum, coated with nickel, polished with a gold coating to $\lambda$/20 rms (for $\lambda = 0.55\mu$m) surface error.

Two laboratory images from the IFS are shown in Figure~\ref{spectraanddots}.  The left panel shows an image created by a tunable laser operating at a single wavelength of 1.33 $\mu$m. The array of dots traces the pattern of the microlenses on the array and reveal the 18.43$^\circ$ rotation of the array.  The right panel of Figure~\ref{spectraanddots} shows the spectra obtained when the instrument is illuminated by a broadband infrared source.

\subsection{Detector System}
The heart of the detector system is a Rockwell (now called Teledyne) HAWAII-2 2048x2048 pixel HgCdTe infrared array operating at 77K. The detector control uses a Generation III infrared array controller designed and built by Astronomical Research Cameras, Inc. (ARC). The interface between the detector controller and the chip itself was constructed and tested at the Institute of Astronomy at Cambridge.  

\subsection{Mechanical Design}
We are required to operate our infrared array at cryogenic temperatures, and hence a cryogenic vacuum-chamber dewar is required. This section describes the features of many of the mechanical aspects of the IFS, all of which were built and tested at the American Museum of Natural History.

\begin{figure}[ht]
\center
\resizebox{1.10\hsize}{!}{\includegraphics[angle=0.]{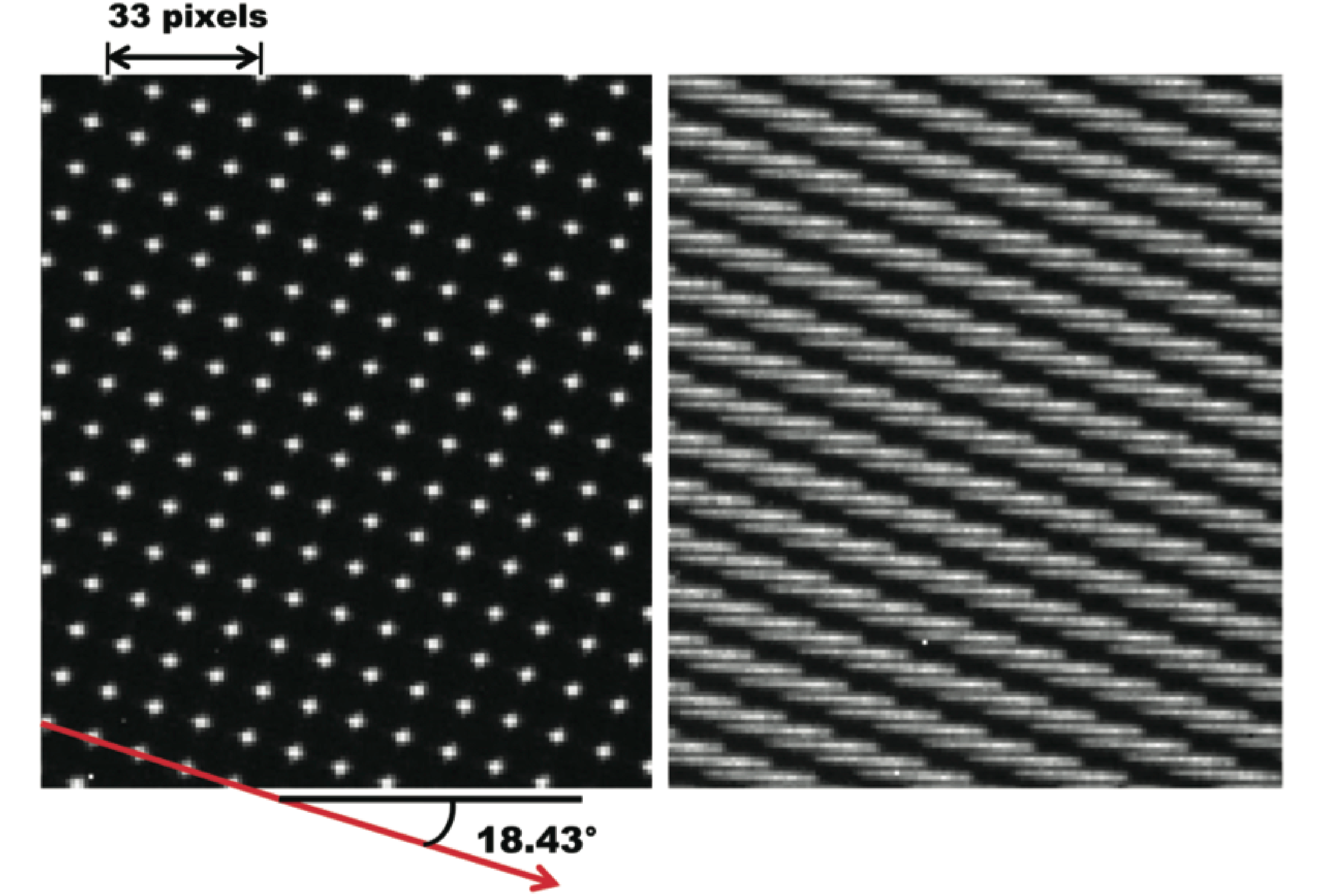}}
  \caption{Laboratory calibration data from the IFS showing a monochromatic 1.33 $\mu$m source (left) and broadband source (right). The orientation of the light pattern traces the pattern of microlenses on the array, which has been rotated by 18.43$^\circ$. }
  \label{spectraanddots} 
\end{figure}

\subsubsection{Dewar}
Our cryogenic dewar is very similar to that used for the PHARO infrared camera at Palomar \citep{hbp01}. Our dewar was built in 2006 by Precision Cryogenics in Indianapolis, Indiana.  The dewar is made almost entirely of 6061-T6 Aluminum with an outer shell divided into an upper and lower part. The lid ranges from $\frac{1}{4}$ to $\frac{3}{4}$ inches in thickness and provides strong support for the overall assembly, while the lower half is lighter weight. The lower portion containing the IFS optics is shown in Figure~\ref{dewarinternals}.  These two halves wrap around the workplate, the inner heat shields, and the two Liquid Nitrogen tanks and each half has a $\frac{3}{4}$-inch flange, or lip, where the two are joined. The optics base plate is comprised of a 1-inch thick, light-weight piece of aluminum and is mounted to the outer portion of the dewar by four mounting tabs made of G-10 fiberglass. These tabs help the optics plate to be thermally insulated from the warm outer portion of the dewar.  

Inside the outer surface of the dewar are the upper and lower radiation shields. The shields are wrapped in multiple layers of thin mylar insulation.  Like the PHARO dewar, the Project 1640 dewar has two separate liquid nitrogen tanks: a 3.3L inner can directly bolted to the optics base plate, and a larger 11L can maintaining close contact with the radiation shield. This larger tank serves as the more global dewar cooler, while the small can provides a local heat sink for the detector and optics. 
The internal parts of the dewar can remain at 77K for 60 hours without refilling the nitrogen tanks.   

The bottom half of the Project 1640 dewar is similar to that for PHARO, but with liquid nitrogen tanks switched. 
The liquid nitrogen fill holes are in the same place as on PHARO, but unlike PHAROs five ports, this dewar has four ports: two for the liquid nitrogen inputs into each can, one for the vacuum pump, and another for attachment of a vacuum gauge.

\subsubsection{Mounting and flexure control}
To maintain a stable dewar position relative to the coronagraph and AO system, and to minimize flexure during telescope slewing, we have have developed a custom dewar handling bracket manufactured by Opticology in New York City. Our dewar has three mounting pins, two towards the front, and one on the rear face, which are used to attach to our mounting bracket. The entire bracket assembly is mounted on Bosch-Rexroth flexure-resistant rails which allows 20-30 mm of focus movement using a fine-thread screw. This mounting bracket also allows a $\pm$10 degree tilt (``pitch'') using a screw-jack mechanism at the rear of the dewar, allowing the entire dewar to pivot on its front two mounting pins by applying a vertical movement at the rear pin. 

\begin{figure*}[ht]
\resizebox{1.05\hsize}{!}{\includegraphics[angle=0.]{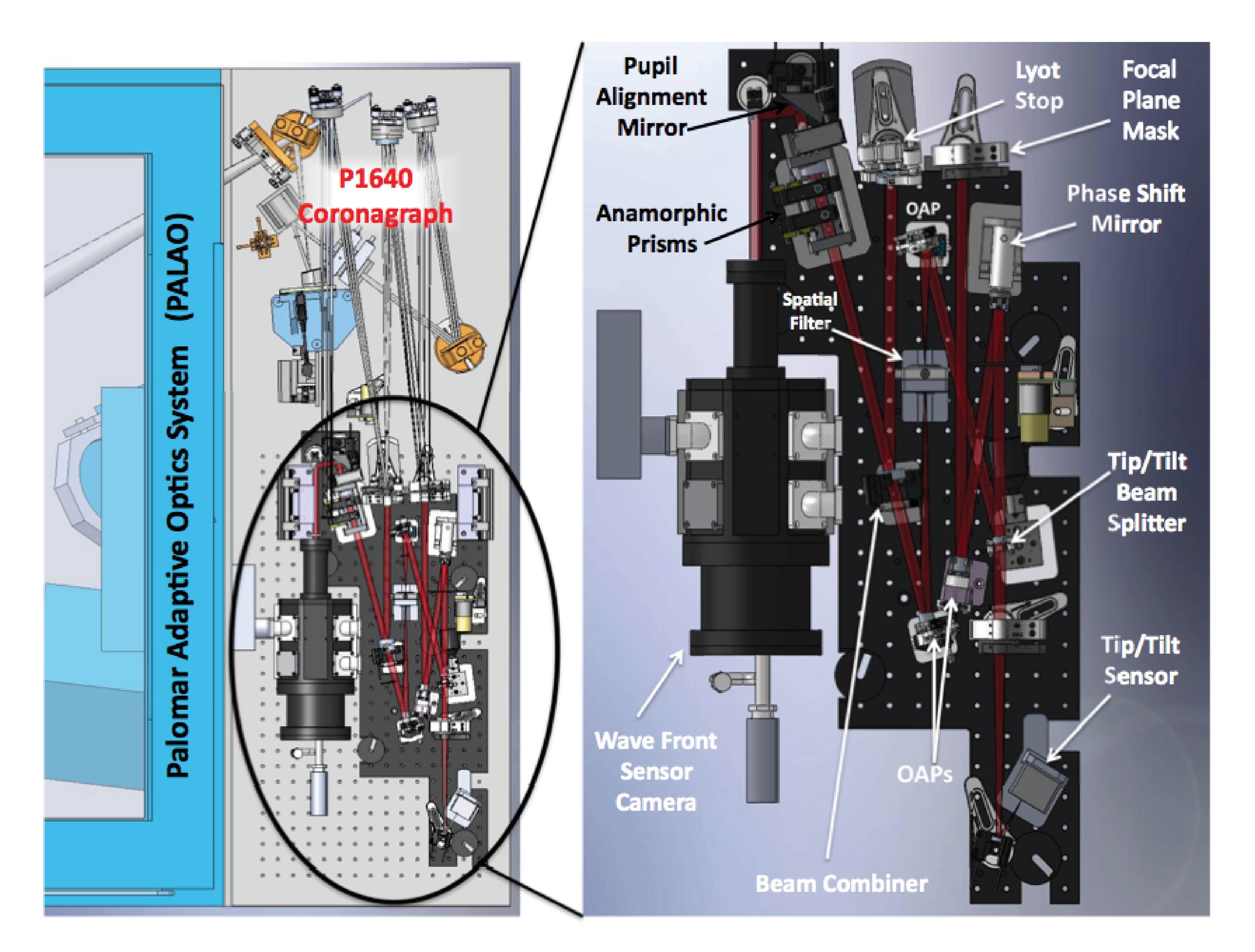}}
  \caption{Details of the post-coronagraph wave front calibration subsystem based on the design of \citet{wgs04,wbr06}.  The left panel shows the full instrument footprint, including the Palomar Adaptive Optics System, and our Apodized Pupil Lyot Coronagraph (Figure~\ref{coronagraph}). The IFS is out of the plane of these optics and is not shown here for clarity. The right panel shows a detailed diagram of the calibration system.   This subsystem actively senses phase aberrations in the coronagraph wave front, and provides centroid offsets to the Deformable Mirror to pre-compensate for these aberrations.  These aberrations are the source of the quasi-static speckle noise, which greatly inhibits our ability to detect faint companions to nearby stars.}
  \label{caldetail} 
\end{figure*}

\subsection{Detector System Software}
The electronics controller for our HAWAII-2 infrared detector was manufactured and designed by ARC \citep{lbe98}. In order to maintain the greatest amount of flexibility and portability, our collaborators at the Astronomical Technology Centre in Edinburgh have configured our detector system to communicate with the observer's control computer using XML files that are transferred using the http protocol \citep{bva02}.  The http protocol was chosen to allow maximum flexibility and stability when such a system is moved from a particular institution or telescope.  The XML files include all of the necessary parameters for a particular observation (exposure time, number of reads, etc). While we use Non-destructive read (NDR) mode for data acquisition, the system can also perform Correlated Double Sampling (CDS). The user sends the appropriate configuration XML files via http to a set of three separate, but connected, servers running on our Data Acquisition Computer that organize the operations of the detector system. The user can directly communicate with the Camera and Filesave servers, but the Filesave server is the only module that will communicate with the de-multiplexing server. 

We communicate directly to our internal servers, motor controllers, temperature controllers and the Palomar telescope control system using customized LabVIEW software.  These servers communicate with the ARC detector controller, which in turn, organizes the reading of the infrared array through the timing and clock boards. When an exposure is complete, the data files are stored in a raw data format, and the de-multiplexing http server converts these into FITS files. 

\section{Palomar AO System}\label{palao}
For the last $\sim$14 years, the Palomar AO system (PALAO) has been based on a 241-actuator AO system built by JPL for use on the 200-in Hale Telescope at Palomar \citep{dwb97,tdb00}.  In mid-2010, the Palomar AO system was removed from the Cassegrain focus of the Hale 200-in telescope and de-commissioned for several months to facilitate an upgrade to a much higher-order corrective system.  This new system, termed ``PALM-3000'' \citep{dbb06,dbr07} aims to achieve extreme-AO correction in the near-IR, as well as diffraction limited imaging in the visible. The primary corrective optic in the new system is a 3388-actuator deformable mirror (hereafter, ``DM''), which will correct the wave front aberrations at high-spatial frequencies (the ``tweeter''). However, the system will also make use of the original 241-actuator DM (the ``woofer'') to correct the low-order aberrations. In addition to a new infrared tip/tilt sensor, the system uses a 64$\times$64 Shack-Hartmann wavefront sensor, with adjustable pupil sampling of 8, 16, 32, and 64 subapertures across the pupil \citep{bar08}.  The coarser pupil sampling allows for guiding on fainter stars.  The wave front control computer is based on a cluster of 17 graphics processing units.  The highest actuator density, sampling the pupil at $\sim$8.1cm, dictates a control radius ($N_{\hbox{act}}\lambda/2D$) of 1.8 arcseconds at 1.78 $\mu$m using Natural Guide Stars (NGS), where $N_{\hbox{act}}$ is the linear number of actuators across the pupil.  This radius is well-matched to the Project 1640 field of view \citep{bda09}.

\subsection{Integration with Palomar Adaptive Optics System}
Our entire coronagraph+IFU package is mounted on a single Thorlabs custom breadboard $18^{\prime\prime} \times 54^{\prime\prime} \times 2.4^{\prime\prime}$ in size. One face of our breadboard has $\frac{1}{4}$-20 tapped holes on a one inch grid, suitable for mounting the coronagraphic optics and the dewar mounting bracket. The other side of this breadboard contains four custom aluminum pucks for mounting the entire assembly to the Palomar AO system. The AO bench has an identical set of pucks attached in the same configuration. When the instrument is raised up to the bench, the four opposing sets of pucks are aligned and clamped together, with the dewar and coronagraphic optics hanging down (Figure~\ref{p1640photo}).  Mounting the instrument in this manner each time ensures that the alignment of the instrument to the AO system is repeated to to less than 1 mm.

\begin{figure*}[ht]
\center
\resizebox{1.25\hsize}{!}{\includegraphics[angle=0]{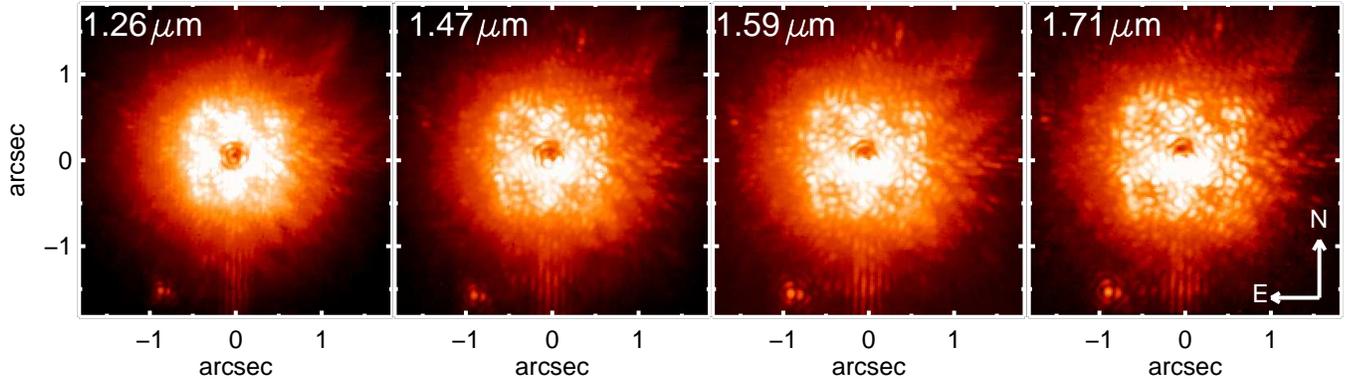}}
 \caption{Four data cube slices at 1.26, 1.47, 1.59, and 1.71 $\mu$m showing a coronagraphic observation from the integral field spectrograph and coronagraph.  The star,  $\zeta$ Virginis \citep{hob10}, has been occulted by our focal plane mask, and its companion is visible near the 7 o'clock position. The evolution of the quasi-static speckle pattern with wavelength is evident in the progression of images, and this wavelength diversity can be used to separate this highly static speckle noise from true astrophysical companions such as the one in this image. }
  \label{zetavir} 
\end{figure*}

We have also developed a customized handling cart for smooth instrument transport.  The cart design aids in the installation on the AO bench primarily via two features: 1) Six spring housings cushion the transport as well as provide differential compression: as the instrument is raised up on the Cassegrain elevator, the slightly uneven elevator floor often causes one portion of the instrument to reach the optical bench first. Compression in this corner will allow the instrument to become parallel with the AO bench as the instrument is raised up. 2) Our cart has fine $x$-$y$ adjustment to match our mounting pucks with the AO bench pucks. The instrument can be rotated on this cart in a spit-like manner---useful for switching between the optics down configuration for mounting, and the optics up configuration for instrument maintenance.

\section{Precision Wave Front Calibration Unit}\label{calsystem}
To achieve a wave front irregularity of $\sim$10 nm rms precision, a customized wave front calibration system has been developed at the Jet Propulsion Laboratory and implemented into the instrument envelope which contains the coronagraph and IFS. This post-coronagraph calibration interferometer subsystem actively senses the internal coronagraph wave front, \citep[see e.g.][]{ssp08}, at the optimized science band pass (1.65$\mu$m), and provides centroid offsets to the DM.  By providing these offsets ``upstream'' to the deformable mirror, the entire optical train is pre-compensated to provide a corrected wave front to eliminate those wave front errors that are not common with the AO wave front sensor.  

In design, this system is similar to that being designed for the Gemini Planet Imager \citep{wbr06,wbp09}, and is based on a Mach-Zender phase-shifting interferometer which  interferes the pupil beam sampled at the location of the Lyot Stop with a reference beam formed by low-pass filtering the light passing through the hole of the focal plane mask. We employ a phase shifting mirror to induce a phase difference between the two arms of the interferometer before re-combining them.  Details of the system and its configuration relative to the coronagraph and Palomar AO system are shown in Figure~\ref{caldetail}. This system will eventually operate at the $\sim$1 Hz update rate, which is easily sufficient given that this subsystem aims to minimize the quasi-static wave front aberrations in the wave front which give rise to speckles with timescales of hundreds of seconds or longer \citep{hos07}. The infrared camera for the wave front calibration system contains a Teledyne Engineering grade PICNIC array, and is cooled with a Polycold Joule-Thompson cold-head + remote compressor.  In addition, our infrared tip/tilt sensor discussed above in section~\ref{coronagraphsec} is integrated with this subsystem, taking the light transmitted through our focal plane mask hole. A more detailed discussion of this subsystem will be given in a future work.

\section{Data Calibration and Pipeline}\label{pipelinesec}
The integral field spectrograph focal plane consists of $4\times10^4$ closely packed, interleaved spectra---one for each microlens in the field of view (Figure~\ref{spectraanddots}, right hand side). In order to inspect and analyze the data, we translate each focal plane image into a cube: the image on the microlens array at 23 channels spanning our wavelength range ($J$ and $H$-bands). Our data pipeline to produce data cubes, fully described in Zimmerman et al (2010b, in preparation), automates this procedure.

First, the pipeline software prepares the detector data by removing the effect of bad pixels, cosmic rays, bias, thermal counts, and variations in pixel sensitivity. To extract the data to a cube, the pipeline must know what position on the focal plane corresponds to a given ($x$, $y$, $\lambda$) combination in a cube. To map this correspondence, we illuminated the integral field spectrograph with a tunable laser (Figure~\ref{spectraanddots}, left hand side). From the sequence of monochromatic images acquired for each channel wavelength, we derived a lookup table that gives the focal plane coordinates for each combination of microlens position and channel. To form the data cube, the pipeline loops through the lookup table and computes a weighted sum of a 3$\times$3 box of pixels centered at each extraction location. The extraction weights themselves are the normalized point spread functions recorded in the monochromatic images. The weighted sum is stored as one pixel value in the cube.

In practice, this procedure is complicated by variation in the alignment between the spectrograph optics and the detector. However, the pipeline is able to account for this using a cross-correlation registration algorithm. Lastly, we correct the flux values in the data cube for the wavelength-dependent transmission of the atmosphere and the instrument optics. To do this, the pipeline applies an array of scale factors to the channel images making up the cube, compensating for the overall spectral response. The end products of the pipeline are data cubes stored in FITS files, ready for spectrophotometry, astrometry, and advanced post-processing techniques like speckle suppression (Section~\ref{specsuppress}).

\begin{figure}[ht]
\center
\resizebox{1.05\hsize}{!}{\includegraphics{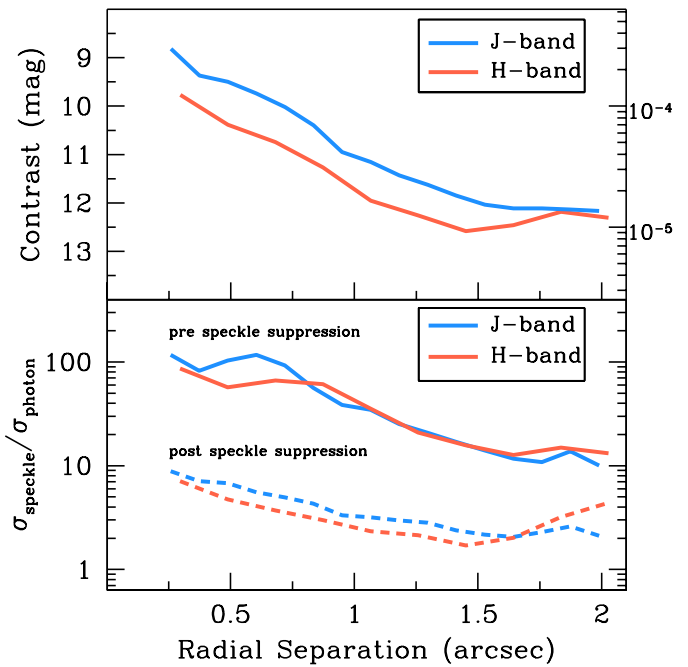}}
  \caption{ {\it Top panel:} Radial $J$ and $H$-band contrast curves, expressed in magnitudes fainter than the host star, for the Project 1640 coronagraph and IFS. Each curve shows the sensitivity measured after applying our speckle suppression algorithm. This contrast incorporates 1200s of exposure time on a bright ($V$=3.9) magnitude A-star under median conditions prior to the installation of the wave front calibration system and high order AO system.  {\it Bottom panel:} A plot showing the $J$ and $H$-band gain in sensitivity after our speckle suppression algorithm (Crepp et al 2010, submitted).  The lines show the rms intensity of the quasi-static speckle noise relative to the photon noise limit.
  The solid curves show the amplitude of the speckle noise relative to the photon noise prior to the application of our speckle suppression algorithm, while the dashed set shows the same after the algorithm has been applied, as shown in Figure~\ref{specklesuppress}.
}
  \label{dynrange} 
\end{figure}

\section{Achieved Performance and Future Directions}
We show an typical example of data acquired with this system in Figure~\ref{zetavir}.  The figure shows four data cube slices at 1.26, 1.47, 1.59, and 1.71 $\mu$m of the star $\zeta$ Virginis \citep{hob10} with the faint stellar companion visible at the lower left of the image.  The evolution of the quasi-static speckle pattern with wavelength is evident in the progression of images, allowing for differentiation between the quasi-static speckles and any true astrophysical companions. 

\subsection{Sensitivity and Achieved Speckle Suppression}\label{specsuppress}
To take advantage of the spectral dependence of the speckle noise pattern, and thereby improving our contrast, we have chosen to base our speckle suppression algorithm on the Locally Optimized Combination of Images  (``LOCI'') method to construct a reference PSF image, which is then subtracted \citep{lmd07}.  The coefficients that dictate the linear combination of images are optimized inside several smaller sub-regions in the image. In the \citet{lmd07} work,  the LOCI algorithm is applied to the Angular Differential Imaging (``ADI'') high-contrast observing mode \citep{mld06}; this same algorithm can be applied to data that are optimized to use spectral deconvolution \citep{sf02}, as is the case for Project 1640.  While the ADI technique utilizes differential rotation between an object fixed on the sky and the speckle pattern, spectral deconvolution utilizes the wavelength dependence of the speckle noise.  We leave the detailed discussion of this technique to future works emphasizing the speckle suppression steps (Crepp et al. 2010, submitted) and also on the accurate extraction of spectra (Pueyo et al. 2011, in prep).

We evaluate our achieved contrast by measuring the local noise amplitude in a subregion of a few $\lambda/D$ in the field of view.  We plot our results in the top panel of Figure~\ref{dynrange}, which is an ensemble average of those channels which encompass the $J$-band (1.06---1.35 $\mu$m) and $H$-band (1.51---1.78 $\mu$m).  The top panel shows the faintest possible source detectable at the 5$\sigma$ level as a function of the radial separation in the field of view. Subsequent to applying our speckle suppression algortihm, we achieve a $H$-band contrast of $\sim$12 magnitudes at 1$^{\prime\prime}$ and $\sim$12.6 magnitudes at 1.5$^{\prime\prime}$. 

\begin{figure}[ht]
\center
\resizebox{1.2\hsize}{!}{\includegraphics{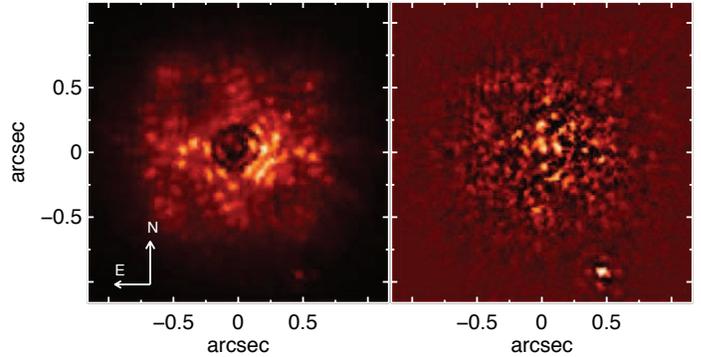}}
  \caption{Two images illustrating the results of the speckle suppression algorithm. A broad-band coronagraphic image of the star Alcor is shown at left with no speckle suppression post-processing. The right panel shows the same star following the application of our speckle suppression algorithm (Crepp et al 2010, submitted).  The faint companion discussed in \citet{zoh10} can be seen in both panels near the lower right in each panel.}
  \label{specklesuppress} 
\end{figure}

The bottom panel of Figure~\ref{dynrange} shows the amplitude of the speckle noise relative to the photon noise, prior to, and following our speckle suppression algorithm.  Plotting the amplitude of the speckle noise relative to the photon noise in this way gives a measure of the amplitude of the speckle noise relative to the fundamental photon noise floor.  {\it Examination of the lower panel of Figure~\ref{dynrange} shows that an improvement of a factor of 10-20 is gained through our speckle suppression algorithm}.  It should be noted at $\sim$1.5$^{\prime\prime}$, we are within a factor of 2-3 of the photon noise limit.  Also, the improvement from 0.25$^{\prime\prime}$ to $\sim$1.6$^{\prime\prime}$ seems to be marginally better for the $H$-band than for $J$, owing to several factors.  Among these are the improved spatial sampling of the point spread function at $H$-band, the improved AO correction at longer wavelengths, and the overall optimization of the system at the $H$-band.  In Figure~\ref{specklesuppress}, we show a typical broadband image along with the same target subsequent to our speckle suppression algorithm.   

Measurements on the first generation PALAO system indicate a remaining RMS uncorrected wave front aberration due to atmospheric effects of $\sim$280 nm.  Error-budget estimates of the new PALM-3000 system predict this residual wave front error will be reduced to 80-90 nm under median seeing conditions, corresponding to Strehl ratios of $\sim$90\%. This expected factor of $\sim$3 reduction in wave front error will translate to an order of magnitude boost in contrast, bringing our current raw $H$-band contrast at 1$^{\prime\prime}$ of $\sim$$2\times10^{-4}$ down to $\sim$$2\times10^{-5}$.  Nevertheless, this 80-90 nm residual wave front error still manifests itself into a smooth seeing halo---a composite of numerous atmospheric speckles averaged together---that can be largely removed through Fourier filtering. However, the Poisson noise on this halo, still a significant limiting factor, can be further suppressed with a $t^{1/2}$ efficiency, where $t$ is the elapsed integration time.  

In addition to this residual atmospheric wave front error, estimates suggest an there is an additional 40 nm of non-common path wave front error intrinsic to the Project 1640 coronagraph and IFS leading to the highly quasi-static population of speckles. Our wave front calibration system will reduce this non-common path wave front error to $\sim$10-15 nm, corresponding to at least an order of magnitude boost in contrast to $\sim$$10^{-6}$ at 1$^{\prime\prime}$.  With an additional order of magnitude improvement from our speckle suppression algorithm as demonstrated in Section~\ref{specsuppress}, we expect our final estimated contrast to approach $10^{-7}$ at 1$^{\prime\prime}$.  

\subsection{Initial Observations and Planned Long-Term Survey}
In the initial phase of observations at Palomar observatory from 2008-10, we have observed $\sim$160 stars, with a heavy focus ($\sim$35\%) on A-stars. G stars ($\sim$25\%), F stars ($\sim$18\%) and K stars ($\sim$13\%) formed the bulk of the remaining sample, while a mixture of B-stars and M dwarfs comprised the rest.  In addition, several solar system objects such as Titan, Io, Uranus and Neptune have been targeted. Initial results, including spectra for the companion zeta Virginis b, shown in Figure~\ref{zetavir} \citep{hob10}, and Figure~\ref{specklesuppress} for Alcor b \citep{zoh10} from this early phase of data taking have been published or are in process.  

We anticipate undertaking a much larger, 100-night survey over five years using the PALM-3000 AO system beginning in 2011. Although the AO system will offer superior correction for $V<8$ stars, the system can correct on fainter targets with a coarser sampling of the pupil than the nominal 63$\times$63 subaperture sampling.  Nonetheless, the primary goal for this portion of the survey will be oriented towards the $\sim$335 stars with $V<8$ and within 25pc.

\acknowledgments
This work was performed in part under contract with the California Institute of 
Technology (Caltech) funded by NASA through the Sagan Fellowship Program.  A portion of this work is supported by the National Science Foundation under Grant Nos. AST-0804417, 0334916, 0215793, and 0520822, as well as  grant NNG05GJ86G from the National Aeronautics and Space Administration under the Terrestrial Planet Finder Foundation Science  Program.  This work has been partially supported by the National Science Foundation Science and Technology Center for Adaptive Optics, managed by the University of California at Santa Cruz under cooperative agreement AST-9876783.
A portion of the research in this paper was carried out at the Jet Propulsion Laboratory, California Institute of Technology, under a contract with the National Aeronautics and Space Administration and was funded by internal Research and Technology Development funds. We are grateful to the efforts of the Derek Ives, Stewart McLay, Andrew Vick, and the other skilled engineers at the UK Astronomy Technology Centre for their configuration of the detector control software.  We also are thankful to Chris Shelton for help in the design of the Atmospheric Dispersion Correcting prisms.  BRO acknowledges Futdi.  Our team is also grateful to the Cordelia Corporation, Hilary  and Ethel Lipsitz, the Vincent Astor Fund, Judy Vale and the Plymouth Hill Foundation.

\bibliography{/Users/shinkley/Desktop/Workspace/papers/MasterBiblio_Sasha}
\bibliographystyle{/Users/shinkley/Desktop/Workspace/mypapers/IFU_PASP_paper/apj.bst}


\end{document}